\documentstyle[12pt]{article}

\def\be{\begin{equation}}
\def\ee{\end{equation}}
\def\ba{\begin{array}{c}}
\def\ea{\end{array}}

\def\ben{$$}
\def\een{$$}
\begin{document}

\titlepage

\begin{center}{\Large \bf Exact solution for Morse oscillator \\
in ${\cal PT}-$symmetric quantum mechanics
 }\end{center}

\vspace{5mm}

\begin{center}

{\bf Miloslav Znojil} \vspace{3mm}

Theory Group,
Nuclear Physics Institute AS CR,
\\CS 250 68 \v{R}e\v{z},
Czech Republic\\

\vspace{5mm}

\end{center}

\section*{Abstract}

${\cal PT}$ invariance of complex potentials $V(x) = [V(-x)]^*$
combines their real symmetry with imaginary antisymmetry. We
describe a new exactly solvable model of this type. Its spectrum
proves real, discrete and ``three-fold", $\varepsilon
=\varepsilon_{n}(j)= (2n+\alpha_j)^2 $, $n = 0, 1,\ldots\ $, with
$\alpha_j\geq 0$ and $j=1,2,3$.

\vspace{9mm}

\noindent
 PACS 03.65.Ge,
03.65.Fd


\newpage


Recently, Daniel Bessis \cite{Bessis} and Bender and Boettcher
\cite{BB} proposed a certain modification of the interpretation of
the one-dimensional Schr\"{o}dinger-like differential equations.
In a way inspired by the methodical relevance of the spatially
symmetric harmonic oscillator in field theory \cite{Itzykson} they
tried to weaken the standard assumption of hermiticity of the
Hamiltonian $H = H^+$, i.e., in the strict mathematical language,
the requirement of the essentially self-adjoint character of this
operator.

The new idea has found its first physical applications and
immediate tests in field theory (cf., e.g., refs. \cite{BM} for
explicit illustrations). Its use in the time-independent quantum
mechanics has simultaneously been proposed \cite{BBjmp}. As long
as the time-reversal operator ${\cal T}$ performs the mere
Hermitian conjugation in the latter case, a generalization of
bound states with definite parity (${\cal P} \psi = \pm \psi \in
L_2$) has been found in normalizable states with a
parity-plus-time-reversal combined symmetry (${\cal PT} \psi = \pm
\psi \in L_2$).

In the ${\cal PT}-$symmetric quantum mechanics a stability
hypothesis $Im \ E = 0$ has been tested semi-classically
\cite{BB,BBjmp}, numerically \cite{Bessis,mytridva}, analytically
\cite{BBjpa,Junker} and perturbatively \cite{mytri,Caliceti}. One
may even return to the related older literature, say, on the cubic
anharmonic oscillator $V(x) =\omega^2+ i g x^3$, all the resonant
energies of which remain real and safely bounded below
\cite{Caliceti,Alvarez}. The similar complete suppression of decay
has now been observed and/or proved in several other analytic
models \cite{Pham,Jones}.

Serious problems arise for the non-analytic ${\cal PT}-$symmetric
interactions \cite{BBjmp}. An indirect clarification of the
difficulty is being sought, first of all, in the various analytic
${\cal PT}-$analogs of the usual square well
\cite{Junker,mytri,SQW}. In the present note, another exactly
solvable example of such a type will be proposed and analyzed in
detail, therefore.

Let us start by characterizing solvable potentials by their so
called shape invariance. The review \cite{Cooper} lists nine of
these models. Once we restrict our attention to the mere
confluent-hypergeometric-type equations on full line, we are left
with the spatially symmetric harmonic oscillator $V^{(HO)}(r)
=\omega^2\left ( r -{b}/{\omega} \right )^2 - {\omega}$ and with
its Morse-oscillator partner
 \be
  V^{(Morse)}(r) =
A^2+B^2\exp(-2\beta r) - 2B(A-\beta/2)\,\exp(-\beta r).
 \label{Mor}
 \ee
The ${\cal PT}-$symmetrization of the former model $V^{(HO)}$ has
been recalled in the pioneering paper \cite{BB}. The latter case
(\ref{Mor}) is to be considered here.

The real and Hermitian version of the model $V^{(Morse)}$ is of a
non-ceasing interest in the current literature \cite{Morhosi}. Its
consequent ${\cal PT}-$symmetrization necessitates a few
preliminary considerations. In the first step, let us make a
detour and return once more to the real and Hermitian $V^{(HO)}$
and to its generalized three-dimensional Schr\"{o}dinger-equation
presentation
\be
\left[-\,\frac{d^2}{dr^2} + \omega^2r^2+
\frac{\ell(\ell+1)}{r^2}\right]\, \psi(r) = E \psi(r), \ \ \ \ \ \
\ \ \ \ell =- \frac{1}{2}+\alpha, \ \alpha > 0.
 \label{SE}
 \ee
We may remind the reader that one can safely return back by
choosing the trivial $\ell = -1$ and/or $\ell = 0$. Still, beyond
this special ``full-line" case with parity $(-1)^{\ell+1}$, the
singularity at $r=0$ remains prohibitively strong for any $\alpha>
0$.

Fortunately, according to our recent letter \cite{PTHO}, an
unexpected remedy may be found in the ${\cal PT}-$symmetrization.
It renders even the singular equation (\ref{SE}) tractable on the
full line. The core of its ${\cal PT}-$regularization lies in an
analytic continuation in the coordinate $r$. This enables us to
work in the complex plane with an appropriately chosen cut. The
cut starts at the singularity in $r=0$ and may be directed,
conveniently, upwards. In such a setting one may integrate
equation (\ref{SE}) along the real line which is only slightly
shifted downwards. In the notation with $r = s - ic, \ c > 0$ and
with a new real quasi-coordinate $s \in (-\infty,\infty)$ the
physical boundary conditions $\psi(-ic\pm\infty) = 0$ remain the
same as before.

A non-numerical key to the complexified one-dimensional problem
(\ref{SE}) lies in a surviving exhaustive solvability of its
differential equation in terms of the confluent hypergeometric
functions,
 \ben
e^{\omega\,r^2/2}\,\psi(x) =  C_1\,r^{-\alpha+1/2}\ _1F_1 \left (
(2-2\alpha-E/\omega)/4, 1-\alpha; \omega\,r^2
 \right )
+
 \een
 \be
+
 C_2\,r^{\alpha+1/2}\ _1F_1 \left (
 (2+2\alpha-E/\omega)/4, 1+\alpha; \omega\,r^2
 \right ).
 \label{newaves}
 \ee
For the large coordinates $r$ which lie within the wedges $|{\rm
Im}\, r| < |{\rm Re}\, r|$, both these hypergeometric series grow
as $\exp(\omega\,r^2/2)$  or, alternatively, degenerate to a
polynomial \cite{Fluegge}. Discarding the former possibility as
manifestly violating our physical boundary conditions we arrive at
a slightly unusual formula for the binding energies
 \ben
 E=E_{qn}=\omega(4n+2 - 2 q \alpha),
 \ \ \ \ \ \ q = \pm 1,\ \ \ \ \ \ \ \ \ n = 0, 1, 2,
\ldots
 \een
and get also wave functions containing Laguerre polynomials,
  \ben
\psi(r) = const. \,r^{-q \alpha+1/2}e^{-\omega\,r^2/2} \ L^{(-q
\alpha)}_n \left [
 \omega\,r^2
 \right ].
 \een
The spectrum remains real, discrete and bounded below and it
reproduces the current, equidistant harmonic energies at $\alpha=
1/2$ \cite{PTHO}.

A change of the coordinates $r = const \times \exp (const\ x)$
mediates a transition to the Morse forces $V^{(Morse)}$. We have
to convert the asymptotically normalizable harmonic wavefunctions
$\psi_{n,\pm}(r) \sim \exp(-\omega\,r^2/2)$, $Re \, r\to\pm\infty$
to their Morse asymptotically normalizable counterparts
$\varphi(x)$. Using the explicit rule
\be
r = -i\,e^{ix}, \ \ \ \ \ \ \psi(r) = \sqrt{r}\, \varphi(x)
\label{chov}
 \ee
in the harmonic Schr\"{o}dinger equation (\ref{SE}) we arrive at
the ${\cal PT}-$symmetrized Morse bound-state problem
\be
\left[-\,\frac{d^2}{dx^2} - \omega^2 \exp(4ix) - D\,\exp(2ix)
\right]\, \varphi(x) = \varepsilon \varphi(x).
 \label{NSE}
 \ee
Our change of variables maps the straight integration contour $r =
s-ic$ with the real $s \in (-\infty,\infty)$ and with the positive
constant $c>0$ onto the deformed, down-bent curve
\be
{\cal C} = \left\{ x = v - iu\ ||\ v \in (-\pi/2,\pi/2),\ u =
u(v)= \ln (c/\cos v) \right \}.
 \label{contu}
 \ee
The circles centered at the origin are mapped upon segments of
horizontal lines. The upward-running cut from $r=0$ to $r=i\infty$
is represented, due to the periodicity of the exponential, by an
infinite family of the vertical lines. The whole plane of $r$ is
mapped on the single vertical strip in $x$, and the multi-sheeted
Riemann surface in $r$ is projected upon the whole complex
$x-$plane.

Our  manifestly ${\cal PT}$ symmetric equation (\ref{NSE})
contains the Morse potential (\ref{Mor}) where $\beta = -2i$,
$B=i\omega$ and $A = i(1-E/2\omega)$. The old energy $E$ is to be
re-interpreted as a new coupling $D=E$. The old centrifugal-like
parameter $\alpha =\ell+1/2$ re-appears as a new momentum and
specifies the Morse energy $\varepsilon = \alpha^2$. The original
harmonic oscillator recipe $\alpha \to E_n$ is replaced by the new
equivalent rule $D \to \alpha^2_m$. This is because the new
integration contour ${\cal C}$ preserves the asymptotic growth or
decrease of the general solutions (\ref{newaves}).

At first sight, the latter conclusion may seem slightly confusing.
Its clarification is easy. In equation (\ref{contu}) the real
parameters $v$ and $u$ are such that the values of $v$ remain
bounded. The variation of $v$ parameterizes not only $s = c\, \tan
v$ but also both the two semi-infinite branches of $u(v)\in (\ln
c, \infty)$. By construction, the harmonic-oscillator states with
the even and odd quasi-parity $q=\pm 1$ and energies $E=E_{n,\pm
1}(\alpha) = \omega(4n+2\mp 2\alpha), \ n = 0, 1, \ldots$ are
mapped upon the new Morse bound states with the energies
 \be
 \varepsilon=\varepsilon_{m}^{[\pm]}(D) = (2m+1\mp D/2\omega)^2,
 \ \ \ \  m = 0, 1,\ldots .
 \label{hor}
 \ee
{\it Mutatis mutandis}, also the wave functions are easily
obtained. One only has to pay attention to the unavoided
\cite{PTHO} crossings at the integers $D/2\omega$.

Phenomenologically, the new spectrum is fairly rich. For
$D/4\omega =M+\sigma-1/2 > 0$ with a small $\sigma \in (0,1)$ and
an appropriate integer $M \geq 0$ the set of the quasi-even
energies splits in the two subsequences. The first one is finite,
 \be
 \varepsilon_{M-k-1}^{[+]}(D)/4 = (k+\sigma)^2,
 \ \ \ \ \ \ \ k = 0,1, \ldots, M -1
 \label{tri}
  \ee
and {\em decreases} with the number of nodes in $\varphi(x)$. The
infinite rest is increasing,
 \be
 \varepsilon_{M+k}^{[+]}(D)/4 = (k+1-\sigma)^2,
 \ \ \ \ \ \ k = 0, 1, \ldots\ .
 \label{dve}
  \ee
In contrast, all the quasi-odd energies exhibit a monotonic growth
and remain minorized by a positive constant $(D/2\omega+1)^2/4$,
 \be
 \varepsilon_{k}^{[-]}(D)/4 = (k+M+\sigma)^2, \ \ \ \ \ \ k = 0, 1,\ldots
 \ ,
 \label{jedna}
 \ee
This split of the spectrum into its three sub-families (\ref{tri})
+ (\ref{dve}) + (\ref{jedna}) is illustrated in Table
\ref{leftvecs} which samples re-orderings of the energy levels in
dependence on the growth of the coupling $D$. Once we ignore the
superscripted quasi-parities (and, perhaps, a possible separation
of the ground state) the spectrum may be visualised as an infinite
sequence of doublets $({\varepsilon_{k}} ,\varepsilon_l)$. Their
square roots are in fact equidistant.

We may summarize that our ${\cal PT}-$symmetrized Morse model
(\ref{NSE}) is completely characterized by its mapping
(\ref{chov}) on the singular harmonic oscillator. One only has to
note that the mapping need not be invertible easily. Indeed, in
the complex $x-$plane, our HO-inspired integration contour ${\cal
C}={\cal C}( -1,1)$ is by far not unique. Firstly, due to the
absence of singularities in eq. (\ref{NSE}) the non-asymptotic
parts of its curved integration contour ${\cal C}$ may be further
deformed. One only has to leave the asymptotics of ${\cal C}$
unchanged. Secondly, an asymptotically different curve ${\cal
C}(-k,l)$ of ref. \cite{Junker} may be chosen as well \cite{Turb}.
It is characterized by the two odd integers $k$ and $l$ and formed
by the respective left and right branches $ x = v - iu $ with $u =
\ln (c/\cos v)$ and $v \in (-k\pi/2,-(k-1)\pi/2)\bigcup
((l-1)\pi/2,l \pi/2)$, pasted together by a (say, straight) line
with $v \in(-(k-1)\pi/2, (l-1)\pi/2)$.

The feasibility of the non-numerical implementation of the latter
idea remains an open question. Formally, the new, more general
Morse-Schr\"{o}dinger differential equation remains ${\cal PT}$
invariant and the manifest ${\cal PT}$ symmetry is also exhibited
by its boundary conditions. In principle, the new spectrum of
energies should remain real and bounded below. In practice, the
proof must be delivered independently. Via a backward mapping, the
modified system would be equivalent to a generalized harmonic
oscillator with the nontrivial integration curve encircling the
essential singularity in the origin and moving (perhaps,
repeatedly, through the cut) to the second (or other) Riemanian
sheet(s).

\section*{Acknowledgement}

Comments on the manuscript and, in particular, on the
supersymmetric background of the Morse oscillator as made by
Rajkumar Roychoudhury from the Physics and Applied Mathematics
Unit of the Indian Statistical Institute in Calcutta during his
recent short stay in \v{R}e\v{z} are gratefully appreciated.

\begin{table}
\caption{ The $D-$dependence and pairing of the low-lying spectra
(\ref{hor}). The level-crossings occur precisely at the even
ratios $D/4\omega$. The energy doublets remain formed in their
vicinity. The equidistance of the moments $
\sqrt{\varepsilon^{[\pm]}_n(D)}$ only takes place at the odd
integers $D/4\omega$. } \label{leftvecs}
\begin{center}
$$
\begin{array}{||c||c|c|c|c|c|c|c|c|c|c|c|c|c|c|c|c|c|c||}
\hline \hline D/4\omega&&   1&&   2&&   3&&
   4&&   5&&   6&&   7&&   8&&
\ldots \\
 \hline\hline
 &\vdots&&   \vdots&&   \vdots&&   \vdots&&   \vdots&&
    \vdots&&
 \vdots&&   \vdots&&   \vdots&\\
 &\varepsilon^{[+]}_3&&   \varepsilon^{[+]}_3&&   \varepsilon^{[-]}_2&&   \varepsilon^{[-]}_2&&
    \varepsilon^{[+]}_4&&   \varepsilon^{[+]}_4&&   \varepsilon^{[-]}_1&&   \varepsilon^{[-]}_1&&   \varepsilon^{[+]}_5&\\
 &-&&   \varepsilon^{[-]}_2&&   \varepsilon^{[+]}_3&&   -&&
    -&&   \varepsilon^{[-]}_1&&   \varepsilon^{[+]}_4&&   -&&   -&\\
 &\varepsilon^{[-]}_2&&   -&&   -&&   \varepsilon^{[+]}_3&&
    \varepsilon^{[-]}_1&&   -&&   -&&   \varepsilon^{[+]}_4&&   \varepsilon^{[-]}_0&\\
 &\varepsilon^{[+]}_2&&   \varepsilon^{[+]}_2&&   \varepsilon^{[-]}_1&&   \varepsilon^{[-]}_1&&
    \varepsilon^{[+]}_3&&   \varepsilon^{[+]}_3&&   \varepsilon^{[-]}_0&&   \varepsilon^{[-]}_0&&   \varepsilon^{[+]}_4&\\
 &-&&   \varepsilon^{[-]}_1&&   \varepsilon^{[+]}_2&&   -&&
    -&& \varepsilon^{[-]}_0&&   \varepsilon^{[+]}_3&&   -&&   -&\\
 &\varepsilon^{[-]}_1&&   -&&   -&&   \varepsilon^{[+]}_2&&
    \varepsilon^{[-]}_0&&   -&&   -&&   \varepsilon^{[+]}_3&&   \varepsilon^{[+]}_0&\\
 &\varepsilon^{[+]}_1&&   \varepsilon^{[+]}_1&&   \varepsilon^{[-]}_0&&   \varepsilon^{[-]}_0&&
    \varepsilon^{[+]}_2&&   \varepsilon^{[+]}_2&&   \varepsilon^{[+]}_0&&   \varepsilon^{[+]}_0&&   \varepsilon^{[+]}_3&\\
 &-&&   \varepsilon^{[-]}_0&&   \varepsilon^{[+]}_1&&   -&&
    -&&   \varepsilon^{[+]}_0&&   \varepsilon^{[+]}_2&&   -&&   -&\\
 &\varepsilon^{[-]}_0&&   -&&   -&&   \varepsilon^{[+]}_1&&
    \varepsilon^{[+]}_0&&   -&&   -&&   \varepsilon^{[+]}_2&&   \varepsilon^{[+]}_1&\\
 &\varepsilon^{[+]}_0&&   \varepsilon^{[+]}_0&&   \varepsilon^{[+]}_0&&   \varepsilon^{[+]}_0&&
    \varepsilon^{[+]}_1&&   \varepsilon^{[+]}_1&&   \varepsilon^{[+]}_1&&   \varepsilon^{[+]}_1&&   \varepsilon^{[+]}_2&\\
 \hline \hline
\end{array}
$$
\end{center}

\end{table}


\newpage
\begin{thebibliography}{99}


\bibitem{Bessis}
D. Bessis, private communication.

\bibitem{BB}
C. M. Bender and S. Boettcher, Phys. Rev. Lett. { 24} (1998) 5243.

\bibitem{Itzykson}
C. Itzykson and J.-M. Drouffe, Statistical Field Theory (Cambridge
University Press, Cambridge, 1989), Vol. 1, Sec. 3.2.3.

\bibitem{BM}
C. M. Bender and K. A. Milton, Phys. Rev. D 55 (1997) R3255 and
 D 57 (1998) 3595 and J. Phys. { A: Math. Gen. 32} (1999) L87;

C. M. Bender, K. A. Milton and Van M. Savage, LANL preprint
 hep-th//9907045.

\bibitem{BBjmp}
C. M. Bender, K. A. Milton and P. N. Meisinger, J. Math. Phys. 40
(1999) 2201.

\bibitem{mytridva}
F. M. Fern\'andez, R. Guardiola, J. Ros and M. Znojil, J. Phys.{
A: Math. Gen. 32} (1999) 3105;

C. M. Bender, G. V. Dunne and P. N. Meisinger,
 Phys. Lett. A 252 (1999) 272;

 M. Znojil, LANL preprint quant-ph/9906029.

\bibitem{BBjpa}
C. M. Bender and S. Boettcher, J. Phys. { A: Math. Gen. 31} (1998)
L273;

M. Znojil, J. Phys.{ A: Math. Gen. 32} (1999) 4563.

\bibitem{Junker}
F. Cannata, G. Junker and J. Trost, Phys. Lett. { A 246} (1998)
 219.

\bibitem{mytri}
F. M. Fern\'andez, R. Guardiola, J. Ros and M. Znojil, J. Phys.{
A: Math. Gen. 31} (1998) 10105.

\bibitem{Caliceti}
E. Calicetti, S. Graffi and M. Maioli, Commun. Math. Phys. 75
(1980) 51;

C. M. Bender and G. V. Dunne, LANL preprint quant-ph/9812039.

\bibitem{Alvarez}
G. Alvarez, J. Phys. A: Math. Gen. 27 (1995) 4589.

\bibitem{Pham}
E. Delabaere and F. Pham, Ann. Phys. 261 (1997) 180 and Phys.
Lett. A 250 (1998) 25 and 29.

\bibitem{Jones}
M. P. Blencowe, H. F. Jones and A. P. Korte, Phys. Rev. D 57
(1998) 5092.

\bibitem{SQW}
C. M. Bender, S. Boettcher, H. F. Jones and Van M. Savage, LANL
preprint quant-ph/9906057.

\bibitem{Cooper}
F. Cooper, A. Khare and U. Sukhatme, Phys. Reports 251 (1995) 267.

\bibitem{Morhosi}
X.-G. Hu and Q.-S. Li, J. Phys. A: Math. Gen. 32 (1999) 139;

H. Konwent et al, J. Phys. A: Math. Gen. 31 (1998) 7541;

H. Taseli, J. Phys. A: Math. Gen. 31 (1998) 779.

\bibitem{PTHO}
M. Znojil, LANL preprint quant-ph/990520 and Phys. Lett. A, to
appear.

\bibitem{Fluegge}
S. Fl\"{u}gge, Practical Quantum Mechanics I (Springer, Berlin,
1971), p. 167.

\bibitem{Turb}
C. M. Bender and A. V. Turbiner, Phys. Lett. { A 173} (1993) 442.

\end{thebibliography}
\end{document}